\documentclass[aps,prb,groupedaddress,nofootinbib,notitlepage,showpacs,floatfix]{revtex4-1}
\usepackage{amsfonts}

\usepackage{graphicx,graphics,epsfig,times,bm,bbm,amssymb,amsmath,amsfonts,mathrsfs}
\usepackage[normalem]{ulem}
\usepackage{wrapfig}
\usepackage{boxedminipage}
\usepackage{setspace}
\usepackage{subfigure}
\usepackage{dsfont}
\usepackage{braket}
\usepackage[pdftex]{color}
\usepackage[pdfstartview=FitH]{hyperref}

\newcommand{\bes} {\begin{subequations}}
\newcommand{\ees} {\end{subequations}}
\newcommand{\bea} {\begin{eqnarray}}
\newcommand{\eea} {\end{eqnarray}}
\newcommand{\beq} {\begin{equation}}
\newcommand{\eeq} {\end{equation}}

\def\>{\rangle}
\def\<{\langle}

\newcommand{\ignore}[1]{}

\newcommand{\reef}[1]{(\ref{#1})}

\begin{document}
\author{Yichen Chang, Tameem Albash and Stephan Haas}
\affiliation{Department of Physics \& Astronomy, University of Southern California, Los Angeles, California 90089, USA}
\title{Quantum Hall States in Graphene from Strain-Induced Nonuniform Magnetic Fields}
\begin{abstract}
We examine strain-induced quantized Landau levels in graphene. Specifically, arc-bend strains are found to cause nonuniform pseudomagnetic fields. Using an effective Dirac model which describes the low-energy physics around the nodal points, we show that several of the key qualitative properties of graphene in a strain-induced pseudomagnetic field are different compared to the case of an externally applied physical magnetic field. We discuss how using different strain strengths allows us to spatially separate the two components of the pseudospinor on the different sublattices of graphene. These results are checked against a tight-binding calculation on the graphene honeycomb lattice, which is found to exhibit all the features described. Furthermore, we find that introducing a Hubbard repulsion on the mean-field level induces a measurable polarization difference between the A and the B sublattices, which provides an independent experimental test of the theory presented here.
\end{abstract}
\maketitle
\section{Introduction}
%

In the absence of other instabilities, electronic materials in a homogeneous external magnetic field are known to form quantized Hall states whose wave functions are localized at the edges of the system. In the usual Hall bar configuration, these quasi-one-dimensional states, typically visualized with semicircular classical trajectories, carry opposite currents along the two edges and are characterized by the bulkÕs nontrivial Chern number.

\begin{figure}[ht] 
   \centering
\includegraphics[width=2.75in]{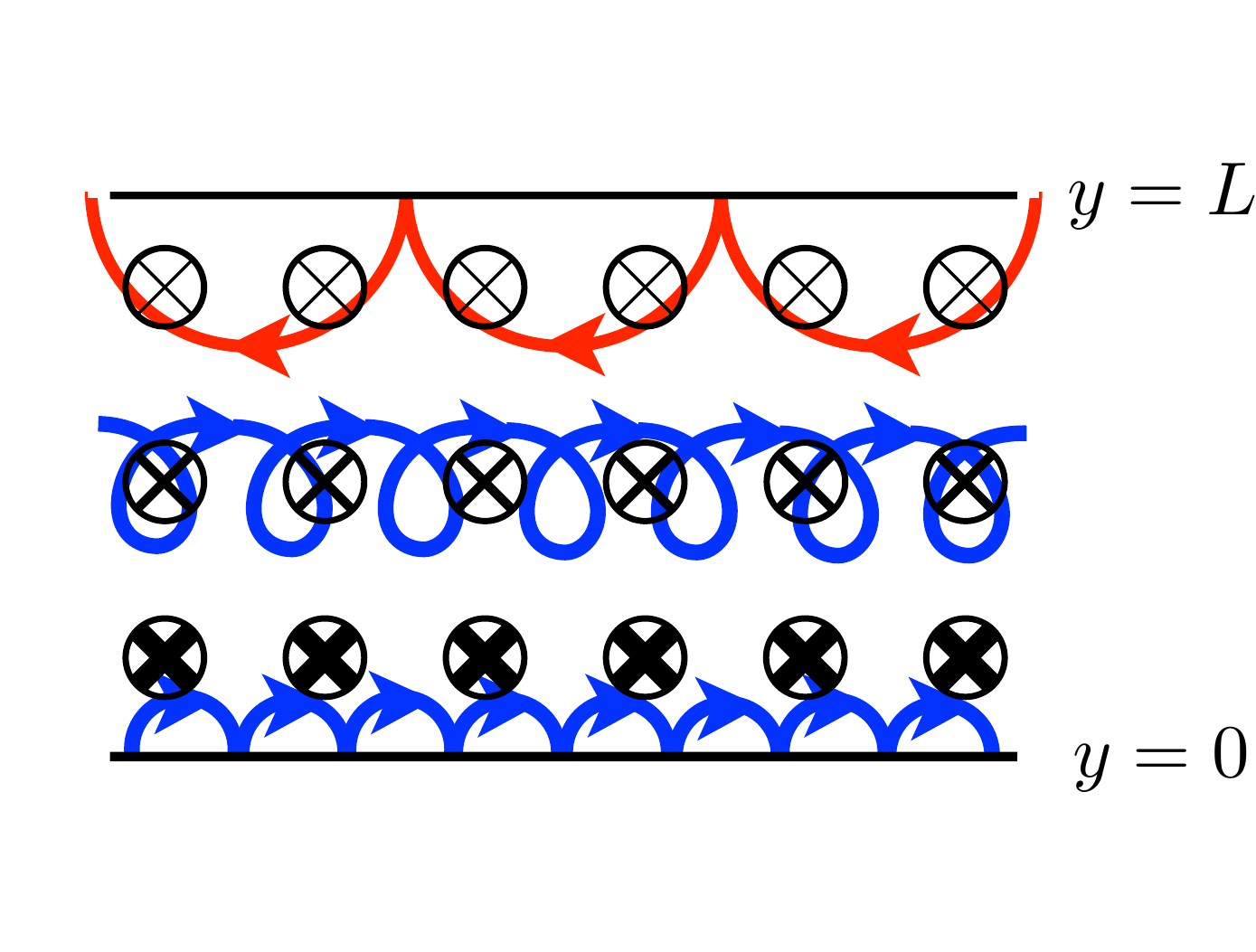} 
   \caption{\small Semiclassical trajectories of the propagating states in a two-dimensional electron gas in a gradient magnetic field. Besides the semicircular edge states, also present in the case of a constant magnetic field, there is a bulk propagating state.} \label{fig:ClassicalOrbits}
\end{figure}

Despite arising from the interaction of noninteracting electrons with a magnetic field, Landau levels can be used to account for many of the features of the integer QHE. For example, the large degeneracy of each Landau level (at large magnetic fields) accounts for the plateaus in resistivity measurements. For a uniform magnetic field, the existence of propagating states only at the edges of the sample can be understood heuristically in terms of the semiclassical trajectories of the electrons near the edges of the sample. Therefore, a characterization of the Landau levels becomes a valuable tool to predict the QHE. Furthermore, the introduction of a gradient in the applied magnetic field can in principle introduce propagating bulk states, \cite{PhysRevLett.68.385} as illustrated in Fig.~\ref{fig:ClassicalOrbits}.

Introducing a strain in graphene quantizes the energy levels,\cite{PhysRevLett.103.046801,Guinea2010} and these quantized energy levels mimic the organization of the Landau levels. For this reason, the language ``pseudo''-magnetic field is used to describe the effect of the strain. In light of the importance of Landau levels to the QHE, strain has been proposed as a useful means to study a strain-induced QHE in graphene as well as controlling the properties of the electrons in graphene. \cite{doi:10.1021/nl1018063}

In this work, we explore the possibility of using strain in graphene to study the Landau level structure in the presence of an effective nonuniform magnetic field. We do this using the Dirac equation coupled to a global $U(1)$ gauge field, which has been shown to be an appropriate effective theory near the $K$ and $K^\prime$ point of graphene.\cite{nature1,nature2}  We focus on the case of a linearly varying magnetic field, starting from a nonzero value on one side of the quantum Hall bar and decreasing in magnitude to 0 on the opposite side. We first show that the only propagating states in the lowest Landau level (LLL) are edge states, and the first instance of chiral modes propagating in the bulk occurs in the first excited Landau level. This is in contrast to the nonrelativistic case \cite{PhysRevLett.68.385} where the LLL contains propagating edge and propagating bulk states. We then demonstrate that for a strain-induced pseudomagnetic field, only a single zigzag edge of the graphene ribbon supports propagating edge modes, unlike the case of an externally applied physical magnetic field that generates oppositely propagating modes at the two zigzag edges. Furthermore, since the two components of the Dirac spinor physically correspond to the wave function on the $A$ and $B$ sublattices of graphene, we find that changing the magnetic field gradient allows us to observe a physical separation between the two spinor components (and hence the wave functions on the $A$ and $B$ sublattices). These results are then confirmed by a tight-binding calculation, where we also show how Hubbard repulsion gives rise to opposite magnetic polarizations in the $A$ and $B$ components of the spinor, thus making their experimental observation using magnetic tunneling tips feasible.
\section{Low-Energy Physics in Terms of Dirac Equations}
We begin by examining the properties of graphene subject to an externally applied physical magnetic field with a finite field gradient, both as a review and to emphasize the differences observed when studying the behavior of graphene in strain-induced pseudomagnetic fields. The low-energy physics near the nodal $K=\frac{2 \pi}{a_0} \left(\frac{1}{3}, \frac{1}{\sqrt{3}} \right)$ and $K^\prime$ $=\frac{2 \pi}{a_0} \left( - \frac{1}{3} , \frac{1}{\sqrt{3}} \right)$  points can be described using a set of Dirac equations, \footnote{Here, we follow the conventions of Ref.~\cite{PhysRevB.73.195408}, although we drop their sign convention around the K$^\prime$ point since we will not be interested in the armchair edge.}:
\begin{equation} \label{eqt:Dirac_RealB}
 (-\sigma_x, -\sigma_y) \cdot \left(  \vec{p} - e \vec{A}  \right) \Psi = \frac{\varepsilon}{v_F} \Psi \ , \quad  (\sigma_x, -\sigma_y) \cdot  \left( \vec{p} - e \vec{A} \right) \Psi' =  \frac{\varepsilon}{v_F} \ \Psi' \ ,
\end{equation}
where $v_F = \gamma_0 a_0 / \hbar$,  $\gamma_0 \approx 3 eV$, $a_0 = 1.42$\AA, and $e < 0$ is the electron charge.  Here we focus on a graphene ribbon geometry extended in the $x$ direction, such that one can use a plane wave ansatz in that direction and decouple the equations for wave functions on the two sublattices at $K$ with $\Psi = (\Phi_A, \Phi_B)\exp(ik_x x)$. This yields a system,
\begin{eqnarray}
& \left(- \partial_y^2 + \left( - k_x + e A_x \right)^2 + e \partial_y A_x \right) \Phi_A = \varepsilon^2 \Phi_A \ ,& \label{eqt:2DEG_EOM1}\\
& \left(- \partial_y^2 + \left( - k_x + e A_x \right)^2 - e \partial_y A_x \right) \Phi_B = \varepsilon^2 \Phi_B \ ,& \label{eqt:2DEG_EOM2}
\end{eqnarray}
and an analogous set of equations at $K^\prime$ with $\Psi' =  (\Phi_A^\prime, \Phi_B^\prime )\exp (i k_x x)$,
where we have chosen the Landau gauge such that only $A_x$ is non-zero.\footnote{We note that the 2nd order equations for $\Phi_A$ and $\Phi_B^\prime$ are identical [giving the equivalent of Eq. \reef{eqt:2DEG_EOM1}] while the equations for $\Phi_B$ and $\Phi_A^\prime$ are identical [giving the equivalent of Eq. \reef{eqt:2DEG_EOM2}].}  There is a unique boundary condition for the pseudospinor components, which is determined by the type of edge termination in the graphene ribbon. \cite{PhysRevB.73.195408}

The most frequently studied termination types for graphene ribbons are armchair and zigzag terminations, although there have been many studies on different types (see e.g. refs.~\cite{PhysRevLett.101.115502,PhysRevB.80.073401,PhysRevB.81.115411,PhysRevB.84.195434}).  At zero magnetic field, armchair terminations do not lead to edge states,\cite{PhysRevB.54.17954} while when a magnetic field is applied, edge states appear at all possible terminations, with the strongest edge states at zigzag terminations \cite{Abanin200777}.   For this reason,  we only consider zigzag terminations at $y = 0$ and $y = L$, leading to the boundary conditions
\begin{equation}
\Phi_A(y = 0) = \Phi_A^\prime (y = 0) = 0 \ , \quad \Phi_B(y = L) = \Phi_B^\prime (y = L) = 0 \ .
\end{equation}
The differential equations for $\Phi_A$ and $\Phi_B^\prime$ can be solved numerically, subject to their respective boundary conditions.  $\Phi_B$ and $\Phi_A^\prime$ are then given by
\begin{equation}
\Phi_B = \frac{-k_x + eA_x - \partial_y}{\varepsilon} \Phi_A \ , \quad \Phi_A^\prime = \frac{k_x - eA_x + \partial_y}{\varepsilon} \Phi_B^\prime \ .
\end{equation}
Assuming a linearly varying magnetic field, we parametrize the gauge field profile as 
\begin{equation} \label{eqt:GaugeField}
A_x = - \frac{\gamma_0}{v_F |e|}  H y \left( \frac{y}{2}  - L \right) \ .
\end{equation}
Furthermore, it is convenient to define dimensionless coordinates and fields given by
\begin{equation} \label{eqt:6}
y= a_0 \tilde{y} \ , \quad L = a_0 \tilde{L} \ , \quad k_x = \frac{1}{a_0} \tilde{k}_x \ , \quad \varepsilon = \gamma_0 \tilde{\varepsilon} \ , \quad A_{x} = \frac{\gamma_0}{ |e| v_F} \tilde{A}_{x} \ .
\end{equation}
In terms of these dimensionless units, the equations of motion are invariant under the scalings
\begin{equation} \label{eqt:8}
\tilde{y} \to \frac{\tilde{y}}{\alpha} \ , \quad \tilde{k}_x \to \alpha \tilde{k}_x \ , \quad \tilde{\varepsilon} \to \alpha  \tilde{\varepsilon}  \ , \quad H \to \frac{H}{\alpha} \ ,
\end{equation}
so the scale invariant quantity one  should consider is the ratio of $H/L$.

\begin{figure}[ht] 
    \centering
    \includegraphics[width=5in]{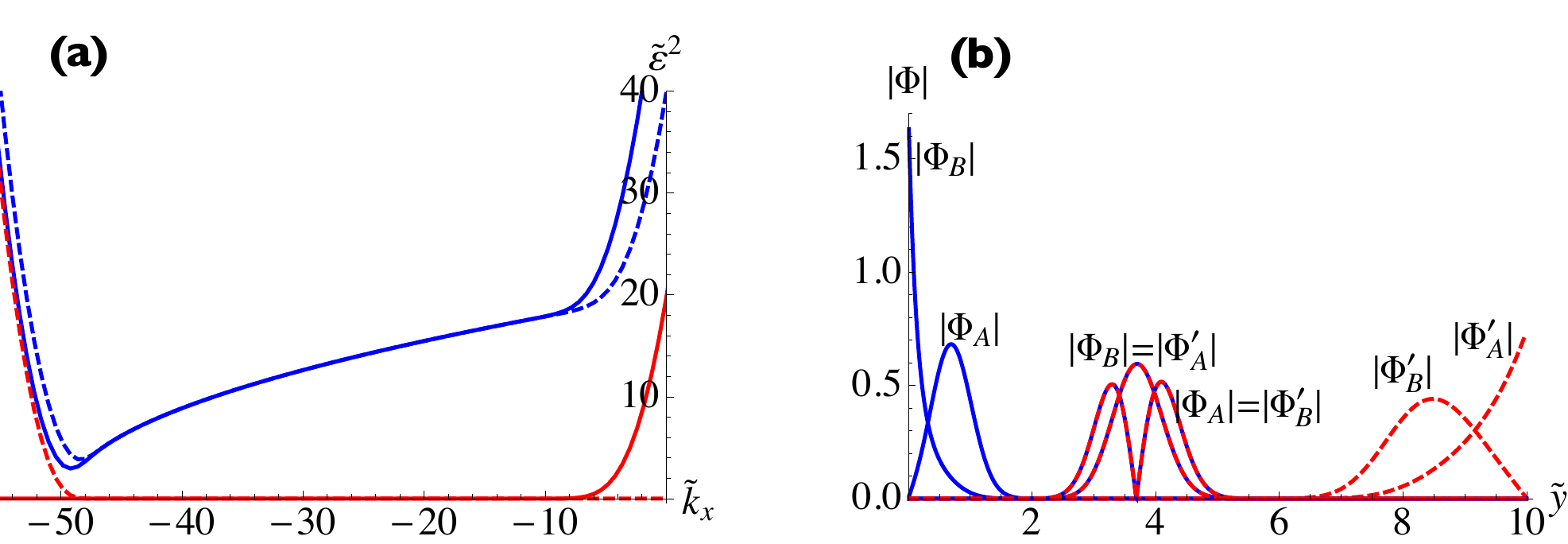} 
   \caption{\small  (a) Energy spectrum around the $K$ and $K^\prime$ points for  $H = 1$ and $\tilde{L} = 10$ [defined in Eq.~\reef{eqt:6}] in the presence of a linearly varying external magnetic field.  The solid (dashed) red curves correspond to the lowest Landau level ($n = 0$) around the $K$ ($K^\prime$) point, and the  solid (dashed) blue curves correspond to the $n=1$ level around the $K$ ($K^\prime$) point.  (b) Magnitudes of the wave functions of the  $n=0$ edge states and $n=1$ bulk state.  The blue curves around $\tilde{y} = 0$ are the LLL evaluated at $\tilde{k}_x = -6.6297$ and the red curves around $\tilde{y} = \tilde{L}$ are the LLL evaluated at $\tilde{k}_x = -48.9882$ corresponding to a dimensionless energy [Eq.~\reef{eqt:8}] of $\tilde{\varepsilon}^2 \approx 0.22$.  The curves around $\tilde{y} = 4$ are $n=1$ bulk states evaluated at $\tilde{k}_x = -30$ around the $K$ point (solid (blue) curves) and $K^\prime$ point [dashed (red) curves] corresponding to a dimensionless energy [Eq.~\reef{eqt:8}] of $\tilde{\varepsilon}^2 \approx 12.61$.}  \label{fig:Graphene_EM}
 \end{figure}

 In Fig \ref{fig:Graphene_EM}(a), we show the energy spectra for the two LLLs. Similarly to the case of a uniform magnetic field, one observes the coexistence of highly dispersive and practically nondispersive regions in momentum space. Note that the spatial variance of the nonuniform magnetic field studied here does not affect this momentum direction. For the LLL($n=0$) around the $K$ and $K^\prime$ points, the dispersionful edge states mimic the behavior observed in the two-dimensional electron gas (2DEG), except that in graphene the $K$ and $K^\prime$ points provide a single edge state each. This is similarly true for the case of a uniform magnetic field, and it is the natural generalization of the single-edge result where only one of the nodal points contributes an edge state \cite{Abanin200777}.  The corresponding spatial profiles of the  wave functions are shown in Fig. \ref{fig:Graphene_EM}(b)\footnote{The normalization of the wave functions at a given energy is given by \cite{PhysRevB.73.195408}
\begin{equation}
\int_0^L d y \left( |\Phi_\mu(y)|^2 + |\Phi_\mu^\prime(y)|^2 \right) = \frac{1}{2} \ , \quad \mu = A, B \ .
\end{equation}
The integral over $|\Phi_A|^2 + |\Phi_A^\prime|^2$ is equal to that over $|\Phi_B|^2 + |\Phi_B^\prime|^2$ for the same energy, so this choice of normalization does not alter the eigenvectors.}.  Around the $K$ point, the exponentially decaying edge state is at $y = 0$ and resides exclusively on the $B$ sublattice, whereas around the $K^\prime$ point, the exponentially decaying edge state is at $y = L$ and resides exclusively on the $A$ sublattice (which is also the case for a uniform magnetic field).  Furthermore, the spatial extent of the edges states at each end of the ribbon is different; the edge state at $y = L$ has a larger spatial extent, which fits nicely with the semi-classical picture of having a larger half orbit at the edge with a lower magnitude magnetic field (which is \emph{not} the case for a uniform magnetic field).  Finally, a propagating bulk mode is  observed in the next higher energy level $(n=1)$, also shown in Fig. \ref{fig:Graphene_EM}(b).  

\begin{figure}[ht] 
    \centering
    \includegraphics[width=5in]{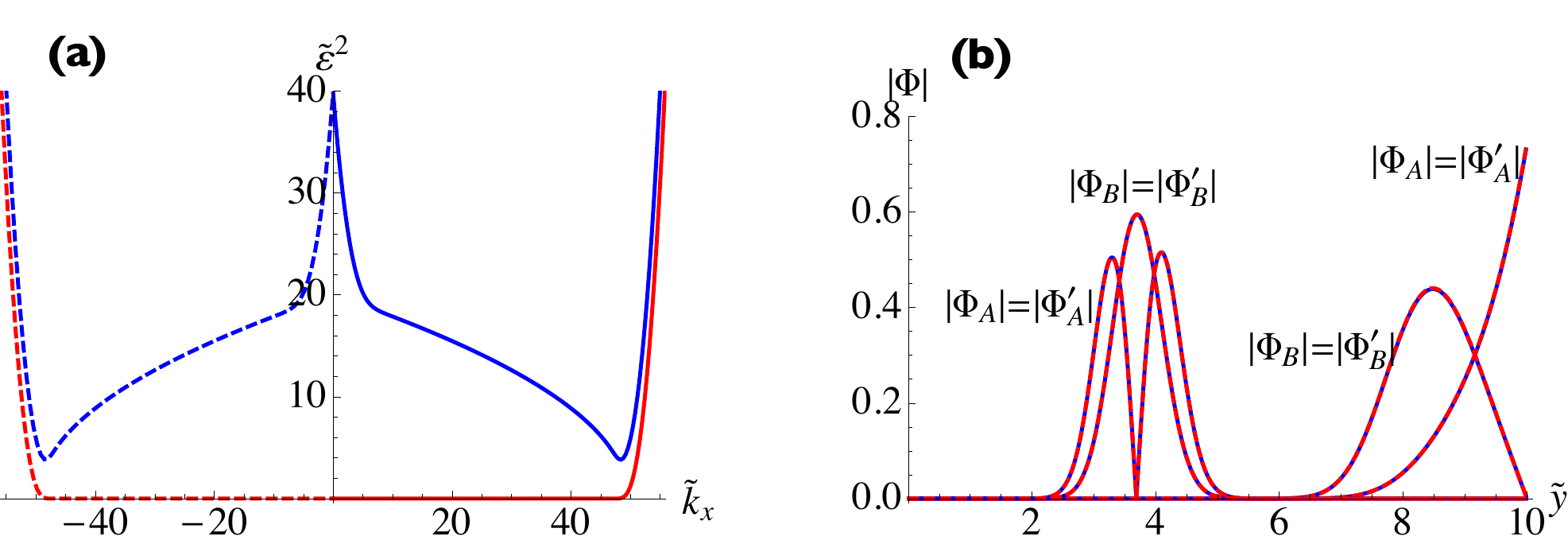} 
   \caption{\small  (a) Energy spectrum around the $K$ and $K^\prime$ points for  $H = 1$, $\tilde{L} = 10$ [defined in Eq. \reef{eqt:6}] in the presence of a strain induced pseudomagnetic field.  The solid (dashed) red curves correspond to the lowest Landau level ($n = 0$) around the $K$ ($K^\prime$) point, and the solid (dashed) blue curves correspond to the $n=1$ level around the $K$ ($K^\prime$) point.  (b) Magnitude of the wave functions of the LLL edge states and $n=1$ bulk state.  The curves around $\tilde{y} = 0$ are the LLL evaluated at $\tilde{k}_x = 48.98707$ (blue curves) and $\tilde{k}_x = -48.98707$ (red curves) corresponding to a dimensionless energy of $\tilde{\varepsilon}^2 \approx 0.22$.  The curves around $\tilde{y} = 4$ are $n=1$ bulk states evaluated at $\tilde{k}_x = -30$ around the $K$ point [solid (blue) curves] and $K^\prime$ point [dashed (red) curves] corresponding to a dimensionless energy [Eq.~\reef{eqt:8}] of $\tilde{\varepsilon}^2 \approx 12.61$.}  \label{fig:Graphene_Strain}
 \end{figure}

We now turn our attention to the effects of a strain-induced pseudomagnetic field in graphene.  The induced pseudogauge field alters the equations of motion at the $K$ and $K^\prime$ points to
\begin{equation}
 (-\sigma_x, -\sigma_y) \cdot \left(  \vec{p} - \vec{A}  \right) \Psi = \frac{\varepsilon}{v_F} \Psi \ , \quad  (\sigma_x, -\sigma_y) \cdot  \left( \vec{p} + \vec{A} \right) \Psi' =  \frac{\varepsilon}{v_F} \ \Psi' \ .
\end{equation}
We assume a similar form for the gauge field profile as in the physical magnetic field:
\begin{equation}
A_x = - \frac{\gamma_0}{v_F} H y \left( \frac{y}{2} - L \right) \ .
\end{equation}
We show in appendix  \ref{app:strain} that for a weak strain as depicted in Fig. \ref{fig:CoordinateSystem}, to a good approximation we can achieve a linear gradient pseudomagnetic field as captured by the above gauge field.  In order to solve for the energy spectrum, we can proceed in a similar fashion as for the physical magnetic field.  Results are presented in Fig.  \ref{fig:Graphene_Strain}(a).  Since the pseudomagnetic field does not break time reversal symmetry, the $K$ and $K^\prime$ spectra are symmetric under sign reversal of $k_x$ in this case.  The spectrum is separated into discrete levels, and the LLL again has a dispersionless regime for a broad range of $k_x$ centered around $k_x  = 0$, which increases in size as the strain is increased.  The $n=1$ level again captures the propagating bulk mode, although the two $K$ points provide oppositely propagating bulk modes.  Focusing our attention on the LLL edge states, the propagating edge states occur for large $|k_x|$ and have opposite group velocities.  In contrast to the case of an externally applied physical magnetic field, we find that the propagating edge states in the pseudomagnetic field exist only at the $y = L$ boundary and solely on the $A$ sublattice [shown in Fig. \ref{fig:Graphene_Strain}(b)].  This is a consequence of the fact that the pseudomagnetic field does not break time-reversal symmetry; since the spectrum allows for only one edge state (propagating in opposite directions) from each nodal point, time-reversal symmetry requires that both modes exist at the same edge.  In order to observe the edge state residing around $y = 0$ (and on the $B$ sublattice), one has to move deep into the dispersionless part of the spectrum.
 As one moves into the dispersionless part of the spectrum, i.e. towards lower values of $|\tilde{k}_x|$, the $\Phi_B$ wave function moves from the $y = L$ edge towards the $y = 0$ edge.   In this process, the $\Phi_A$ wave function grows sharper at the $y = L$ edge.

\begin{figure}[ht] 
    \centering
    \includegraphics[width=5in]{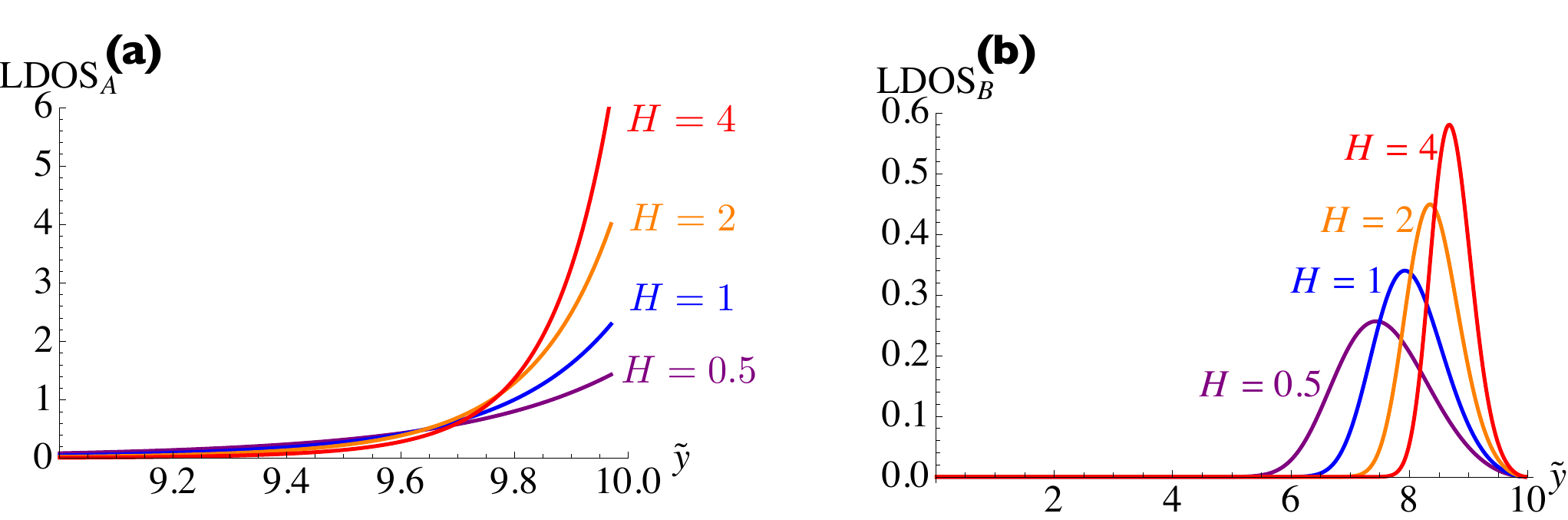} 
 \caption{\small Contributions to the local density of states from the $K$ and $K^\prime$ point states with $\tilde{\varepsilon}^2 \in (0, 0.25]$ for four different gradient pseudomagnetic fields.  (a) LDOS on the $A$ sublattice.  (b) LDOS on the $B$ sublattice.}  \label{fig:Graphene_LDOS}
 \end{figure}

The behavior of $\Phi_B$ as we move into the dispersionless regime results in a noticeable effect on the local density of state (LDOS)\footnote{The LDOS is calculated by summing the magnitude squared of the normalized wave functions from the K and K$^\prime$ point at a given energy.} as the strain is changed.   To illustrate this point, we consider all  states in an energy window $\tilde{\varepsilon} \in [0, \epsilon]$.  We know already that for $\varepsilon \approx 0$, the $\Phi_A$ wave function is strongly peaked around $y = L$ (Fig. \ref{fig:Graphene_LDOS}(a)), and for sufficiently small $k_x$, the $\Phi_B$ wave function is strongly peaked around $y = 0$ .  These states give rise  to an enhanced LDOS at the edges.  Let us now examine the contributions to the LDOS due to the states above zero energy, i.e. the dispersionful states. As shown in Fig. \ref{fig:Graphene_LDOS}(b),  when the field gradient is increased, the peak of this LDOS shifts towards the $y = L$ edge and grows in strength.  Therefore, for sufficiently strong strains this feature manifests itself as a distinct peak away from the $A$ sublattice edge state.
%
\section{Lattice Model Calculations}
%
In order to test whether the effective low-energy Dirac model accurately describes the physics of graphene under strain, we now turn to the Hubbard model on the honeycomb lattice, which has been extensively used to study the magnetic response of graphene (see e.g. refs.~\cite{PhysRevB.77.075430,PhysRevLett.102.136810,PhysRevB.84.115406}), to scrutinize the above results.  Specifically, we model the electronic properties of the graphene sheet using the Hamiltonian
\begin{equation} \label{eqt:Hubbard}
H=\sum_{<i,j,\sigma>}t_{pp\pi}(c{_{i\sigma}^{\dagger}c_{j\sigma} + \mathrm{h.c.})}+U\sum_{i=1}^{N}n_{i\sigma}n_{i-\sigma} \ ,
\end{equation}
where the first term is the single-orbital tight-binding model for graphene (with only nearest-neighbor hopping), and the second term is the on-site Coulomb repulsion between electrons of opposite spin. $c_{i\sigma}$ and $c_{i\sigma}^{\dagger}$ are electron annihilation and creation  operators respectively, $t_{pp\pi}$ is the hopping integral, and $n_{i\sigma}$ is the number operator given by $c_{i\sigma}^{\dagger}c_{i\sigma}$.  
The coefficient $U$ is chosen to be either zero (i.e. a pure tight binding model) or $U=1.2 t_{pp\pi}$, which is below the critical value of $U_{cr}/t_{pp\pi}=2.2$ at which there is a quantum phase transition to an antiferromagnetic insulating phase\cite{0295-5075-19-8-007,springerlink:10.1007/s002570050384}.  For these values of the Hubbard repulsion, the system is in a semi-metallic phase with a conical dispersion, as is experimentally observed in graphene \cite{Novoselov2007}. 
In the case of a finite Hubbard repulsion, the interaction term is simplified using a self-consistent mean field approximation\footnote{This approximation has been found to be valid for the small values of $U/t$ we are interested in for our study \cite{Feldner:2010}.}.  More details of this procedure can be found in appendix \ref{app:MeanField}.

\begin{figure}[ht] 
   \centering
   \includegraphics[width=5in]{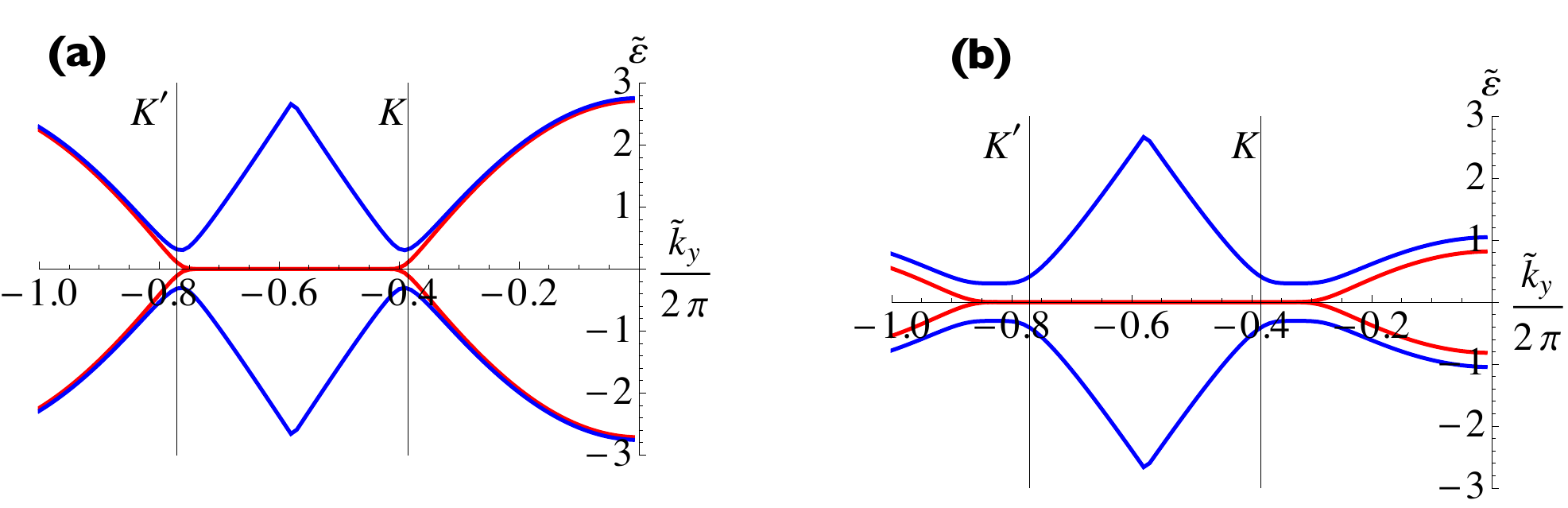} 
 \caption{\small The two lowest-lying energy levels calculated using the tight-binding model ($U = 0$).  (a) Energy spectrum of the tight-binding model on the graphene lattice without a strain.  (b) Energy spectrum of the tight-binding model on the graphene lattice in the presence of an arc-bend strain with  $R=5W$. }  \label{fig:TB_Spectrum}
\end{figure}

Applying a strain to the graphene sheet displaces the  carbon atoms from their equilibrium positions,  and the resulting change in bond lengths can be calculated using \cite{Neto09},
\begin{eqnarray} \label{eqt:bond_length_change}
\delta_a = - \frac{1}{a_0} \vec{R}_a \cdot \left( \vec{u}_i - \vec{u}_{i+a} \right) \ , \quad a = 1,2, 3 \ ,
\end{eqnarray}
where $\vec{u}_i$ is the displacement from equilibrium of the $i$-th $A$ atom at position $(x,y)$ and $\vec{u}_{i+a}$ is the displacement from equilibrium of the neighboring three $B$ atoms.  In turn, the change in the bond length modifies the hopping integral $t_{pp\pi}^{a}$ to\cite{1367-2630-11-11-115002}
\begin{equation}
t_{pp\pi}^{a}= - \gamma_{0}e^{-3.37(\delta_{a})}.
\end{equation}
In order to generate an approximately  linear gradient pseudomagnetic field, as discussed in the previous section, we  consider the very particular inhomogeneous arc-bend strain  \cite{PhysRevB.81.035408} illustrated in Fig. \ref{fig:CoordinateSystem} (here the setup is rotated by 90$^\circ$).  More details can be found in appendix \ref{app:strain}.
The numerical calculations discussed below are performed in real space on a finite 4800-site honeycomb lattice. When the system is finite in the $y$ direction, we find no significant spatial variation in that direction for the lowest energy states in which we are interested, and therefore, to compute the spectrum as a function of momentum we use periodic boundary conditions in the $y$ direction.
\begin{figure}[h]
   \centering
   \includegraphics[width=4in]{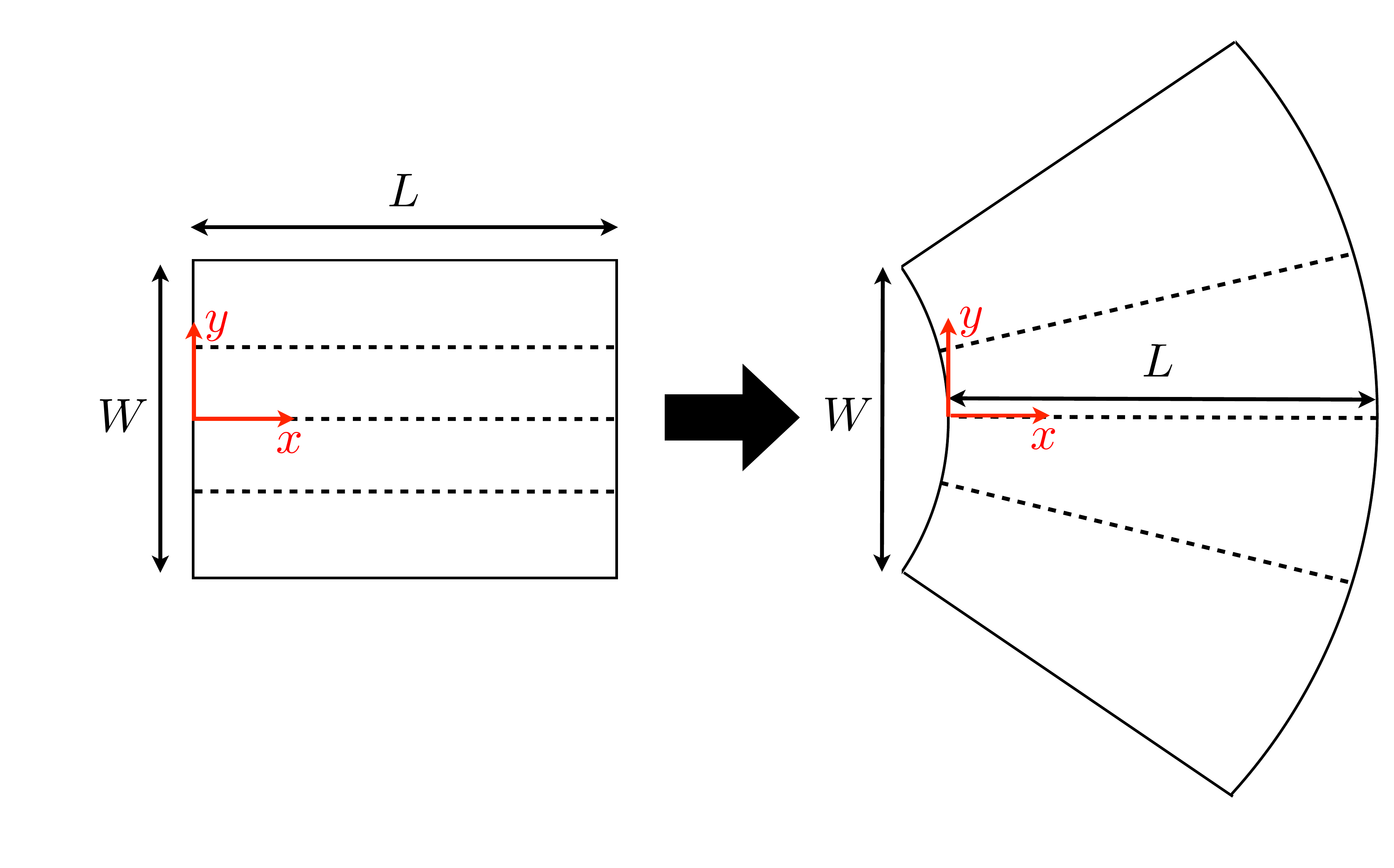} 
   \caption{\small Illustration of an inhomogeneous arc-bend strain, leading to an approximately constant gradient in the induced pseudomagnetic field. An 
originally rectangular graphene ribbon of length $L$ and width $W$ is distorted into a fan-shaped segment of a circular shell with inner radius $R$ and outer radius $R+L$.}
   \label{fig:CoordinateSystem}
\end{figure}
In Fig.  \ref{fig:TB_Spectrum}, the energy spectra of the strained and unstrained graphene  lattice are compared.  In the absence of an applied strain, the spectrum exhibits a dispersionless band between the two nodal points and a subsequent rise in the dispersion near the nodal points that give rise to edge states at zigzag edges\cite{Abanin200777}.  Upon application of the strain, the dispersionless part of the spectrum grows and the behavior around the K and K$^\prime$ points matches the qualitative predictions of the effective low-energy Dirac model discussed above.

\begin{figure}[ht] 
   \centering
   \includegraphics[width=5in]{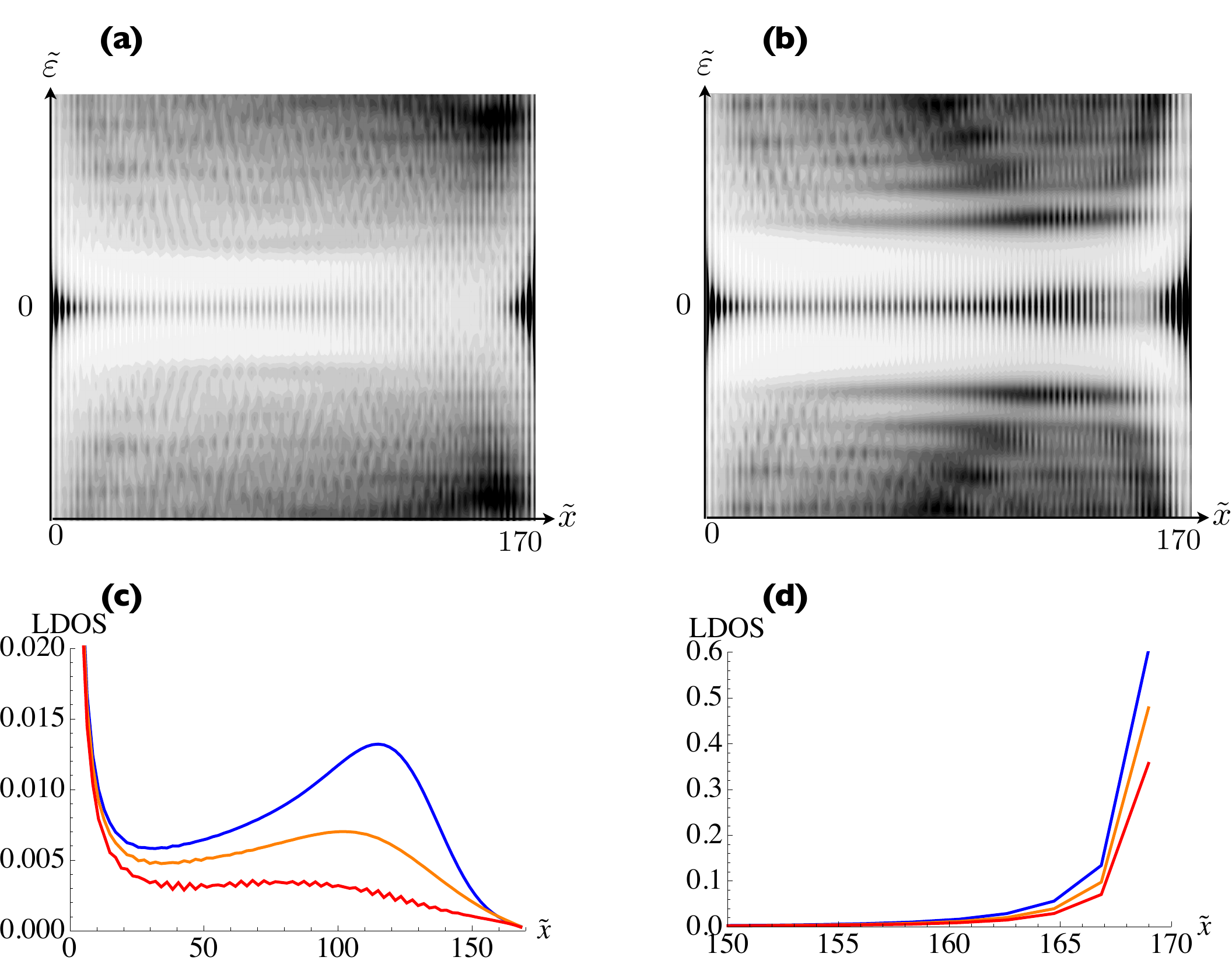} 
  \caption{\small The local density of states (LDOS) as a function of energy in a graphene sheet with arc-bend strain, calculated using the tight-binding model ($U=0$).  (a ,b) Darker shades correspond to a higher intensity LDOS.  (c, d), LDOS near the edges in the tight-binding mode of a graphene ribbon is shown.  The blue curve is for an arc-bend strain of $R = 5.1 W$, the orange curve corresponds to  $R = 3.6 W$, and the red curve corresponds to $R = 3 W$.(a) Arc-bend strain with $R = 5 W$.  (b) Arc-bend strain with $R = 3 W$.  (c) LDOS near smaller inner edge integrated over a small energy window around $\tilde{\varepsilon} = 0$, (d) LDOS near stretched outer edge integrated over a small energy window around $\tilde{\varepsilon} = 0$.}
  \label{fig:TB_LDOS}
\end{figure}

In Fig.  \ref{fig:TB_LDOS}, we show the LDOS as a function of position and energy.  It is observed that for small strains, the LDOS remains peaked at the edges, but as the strain is increased, a bulk LDOS begins to emerge.  On closer analysis (Fig. \ref{fig:TB_LDOS}(c-d), we can see that this bulk LDOS increases  and moves towards the stretched edge as the strain is increased.  Furthermore, the LDOS at the stretched edge grows in intensity while the LDOS at the smaller  edge decreases.  These observations are consistent with the low-energy continuum model. 

\begin{figure}[ht]
   \centering
   \includegraphics[width=5in]{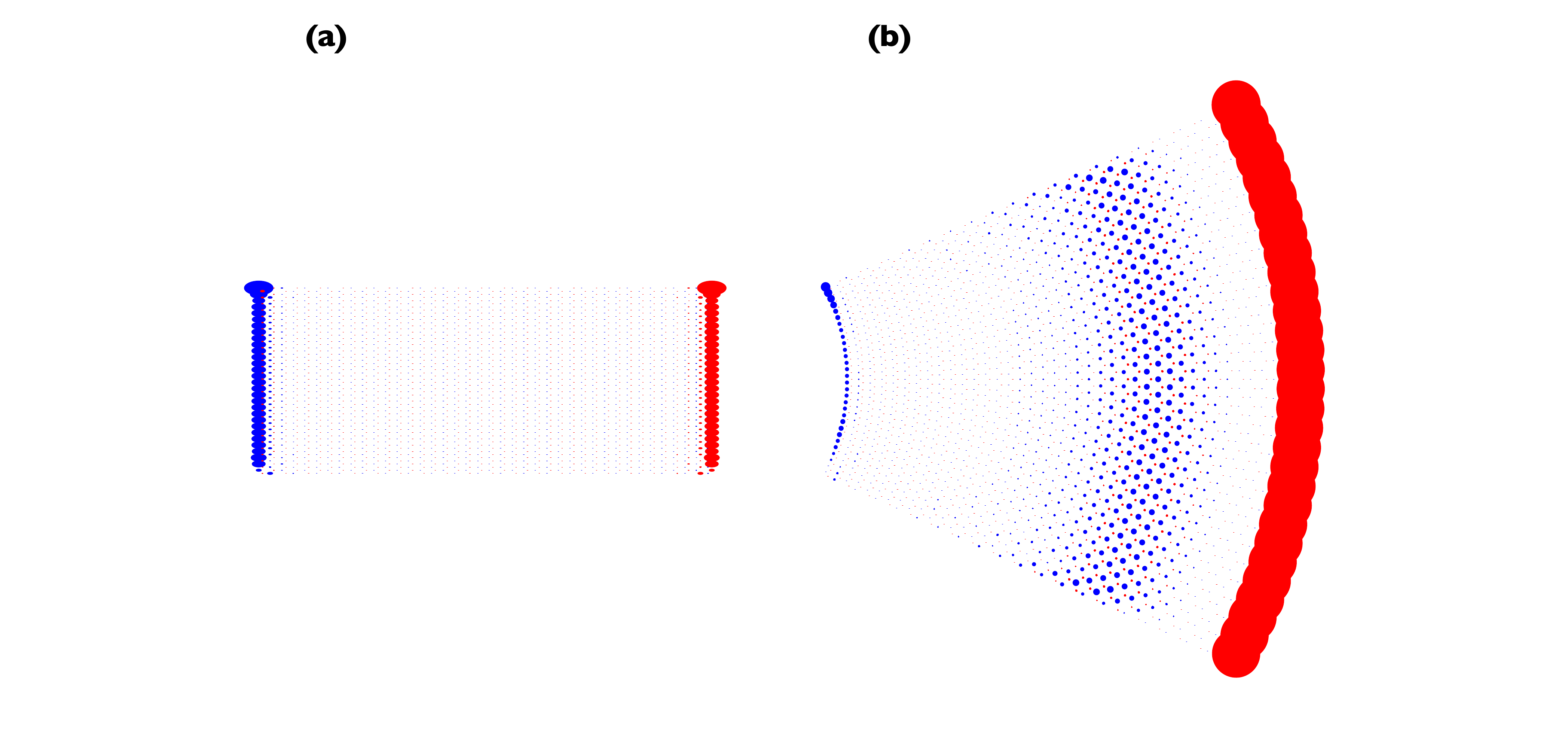} 
   \caption{\small  Local magnetization in a graphene sheet calculated using the Hubbard model.  Red symbols denote negative $z$ polarization, whereas blue symbols denote positive $z$ polarization. (a) Unstrained case.  (b) Strained case.  } \label{fig:TB_LM}
\end{figure}

 Finally, to show that indeed a bulk peak in the LDOS exists on the $B$ sublattice, we consider the case of a finite  Hubbard repulsion.  We find that turning on $U/t>0$ does not change the features of the  results so far, except that the system now exhibits a local polarization.  We show the local polarization in Fig.  \ref{fig:TB_LM}.  Here it is observed that the stretched outer edge is associated with a negative polarization, whereas the smaller inner edge and the bulk mode are associated with a positive polarization, demonstrating that indeed the smaller edge and bulk mode exist on the same sublattice.  

As observed in the effective low-energy continuum model, the bulk mode and the stretched edge mode are actually part of the same dispersionful solution of the Dirac equation, and it is therefore very interesting to see such a distinct separation between the two.
   %
%
   %
%
\section{Conclusions}

We have discussed how strain-induced non-uniform pseudomagnetic fields in graphene  lead to propagating quantum Hall modes with 
non-trivial properties. In particular, we considered the case of an approximately linearly varying pseudomagnetic field, caused by an arc-bend strain. In contrast to the case of an externally applied magnetic field, the propagating pseudomagnetic field edge modes are predominantly localized on the stretched outer zigzag edge of the graphene ribbon.  In addition, we observe that we can spatially resolve the $A$ and $B$ components of the pseudospinor for this propagating mode with increasing arc-bend strain, whereby the $B$-component appears as a peak in the bulk (as opposed to the edge as in the $A$ case).  We also find that a finite Hubbard repulsion leads to a polarization of the propagating modes, which makes them accessible to spatially resolved magnetic measurements.

In order to evaluate the feasibility of observing this phenomenology experimentally, let us now discuss  some estimates. In the arc-bend scenario we considered,  the strain percentage is given by $W/2R$, and the maximum strain occurs at the outer edge.
For example, if $R=3W$, the maximum strain is about 17\%, and for $R=5W$, it would be 10\%.
Previous theoretical papers have predicted that strains in graphene  up to 20\% can be achieved \cite{ PhysRevB.76.064120}, whereas so far 
experimental papers have reported  strains up to 12\% \cite{Lee18072008}. Therefore, the regime of strains that were explored in our model 
calculations  are of the same order of magnitude of what can be realized experimentally. 
Furthermore, we expect that a similar phenomenology, i.e. strain induced polarized edge and bulk modes, should be observable for other strain 
geometries which may be easier to realize, such as local deformation of the lattice. In this context, we would be particularly interested in engineering
strain profiles that allow for both positive and negative induced pseudomagnetic fields, which would permit us to study the snake mode 
scenario which had been proposed \cite{PhysRevB.69.153304,PhysRevB.77.081403,PhysRevLett.107.046602}.  Furthermore, the strain-induced vector potential we have studied here accounts for the hopping perturbation only.  There are corrections to the vector potential arising from deformations of the Brillouin zone that are not the same at the K and K$^\prime$ points \cite{PhysRevB.85.115432}, and it would be interesting to study how the physics discussed here is corrected by these terms.

\section*{Acknowledgements}
We wish to thank D. Ben-Zion, C.N. Lau,  E. Rezayi and H. Saleur for useful discussions, and we gratefully acknowledge financial support by the Department of Energy, Grant
No. DE-FG03-01ER45908.
The numerical computations were carried out on the
University of Southern California high performance supercomputer cluster.
\appendix
\section{Self-Consistent Mean Field Approximation} \label{app:MeanField}
Within  mean field theory, the repulsion term in the Hubbard model  \reef{eqt:Hubbard} can be approximated by 
\begin{equation}
U\sum_{i}^{N}n_{i\sigma}n_{i-\sigma}  = U\sum_{i}^{N} \left(-\left\langle n_{i\sigma}\right\rangle \left\langle
n_{i-\sigma}\right\rangle +\left\langle n_{i\sigma}\right\rangle n_{i-\sigma } +  n_{i\sigma} \left\langle n_{i-\sigma } \right\rangle \right) \ ,
\end{equation}
where the mean field  $\left\langle n_{i\sigma }\right\rangle $ is computed self-consistently from
\begin{equation}
\left\langle n_{i\sigma}\right\rangle =\int dEg_{i\sigma}(E)f(E-E_{f}) \ .
\end{equation}
Here $g_{i\sigma}(E)=\sum_{j}\Psi_{i}^{\ast}(E_{j})\Psi_{i}(E_{j} )\delta(E-E_{j})$ is the local electronic density of states, $E_j$ is the $j$-th energy eigenvalue, and $f(E-E_{f})$ is the Fermi function. The self-consistent solution provides the local densities of states and the spin densities $M_{i}$  on each lattice site, given by
\begin{equation}
M_{i}=(\left\langle n_{i\sigma}\right\rangle -\left\langle n_{i-\sigma}\right\rangle )/2 \ .
\end{equation}
%
\section{Strain-Induced Pseudomagnetic Field} \label{app:strain}
Deforming each lattice point in graphene from position $(x,y)$ to position $(x+u_x, y + u_y)$ leads to an   arc-bend strain  \cite{PhysRevB.81.035408},
\begin{eqnarray}
x + u_x &=& \left( R + x - \frac{W}{2} \right) \cos \theta(y) - R + \frac{L}{2} \ , \nonumber \\
y + u_y &=& \left( R + x - \frac{W}{2} \right) \sin \theta(y)  \ ,
\end{eqnarray}
where $\theta(y) = 2 y \arcsin(W/2R)/W$, and $R$ is the radius of the inner side.  This induces a gauge field given by  \cite{PhysRevB.81.035408}
\begin{equation}
A_x =   -2 c \frac{\beta}{a_0} \partial_y u_x \ , \quad A_y = c \frac{\beta}{a_0} \left( \partial_x u_x - \partial_y u_y \right) \ ,
\end{equation}
where $\beta = - \partial \ln(\gamma_0) / \partial \ln(a_0) \approx 2$.
The resulting pseudomagnetic field in the $z$ direction is then given by
\begin{equation}
B_z =  - \frac{2 c \beta}{W a_0} \cos \theta(y) \arcsin \left( \frac{W}{2 R} \right) \left( 1 - \frac{4}{W} \left( R+ x - \frac{L}{2} \right) \arcsin\left( \frac{W}{2 R} \right) \right)
\end{equation}
In the limit of weak strain ($W/R \to 0$), the pseudomagnetic field is a constant,
\begin{equation}
\lim_{W/R \to 0} B_z = - \frac{3 c \beta}{R a_0},
\end{equation}
whereas for finite but weak strains the induced pseudomagnetic field is approximately linear in $x$ with maximum magnitude at $x =0$ and minimum magnitude at $x =L$.  
\bibliography{ref}

\end{document}